\documentclass[aps, 10pt, prl,
               notitlepage, twocolumn,
               superscriptaddress, longbibliography,
               floatfix]{revtex4-1}

\usepackage{float}
\usepackage{placeins}

\usepackage{amsmath}
\usepackage{amssymb}
\usepackage{amsfonts}
\usepackage[utf8]{inputenc}
\usepackage[T1]{fontenc}
\usepackage{mathrsfs}
\usepackage{anyfontsize}
\usepackage{xspace}
\usepackage[inline]{enumitem}

\usepackage[linktocpage,breaklinks]{hyperref}
\usepackage[usenames,dvipsnames]{xcolor}

\usepackage{tensor}
\usepackage{bm}
\usepackage{dsfont}

\usepackage{graphicx}
\usepackage{epsfig}
\usepackage{epstopdf}

\hypersetup{colorlinks=true,
            citecolor=NavyBlue,
            linkcolor=NavyBlue,
            urlcolor=NavyBlue}

\newcommand{\bfi}[1]{\textit{#1$-$}}

\def\canuda{\textsc{Canuda}}
\def\ETK{\textsc{Einstein Toolkit}}
\def\newacronym#1#2#3{\gdef#1{\gdef#1{#2\xspace}#3 (#2)\xspace}}
\newacronym{\amr}{AMR}{adaptive mesh refinement}
\newacronym{\bh}{BH}{black hole}
\newacronym{\bhs}{BHs}{black holes}
\def\bssn{Baumgarte-Shapiro-Shibata-Nakamura\xspace}
\newacronym{\eft}{EFT}{effective field theory}
\newacronym{\eob}{EOB}{effective-one-body}
\newacronym{\gr}{GR}{General Relativity}
\newacronym{\gw}{GW}{gravitational wave}
\newacronym{\gws}{GWs}{gravitational waves}
\newacronym{\hpc}{HPC}{High-performance computing}
\newacronym{\sGB}{sGB}{scalar Gauss--Bonnet}

\def\dif{\textrm{d}}

\def\aGB{\alpha_{\rm GB}}
\def\RGB{\mathscr{G}}

\def\R4D{R}
\def\Kphi{K_{\rm \Phi}}
\def\rex{r_{\rm ex}}
\def\mmode{\mathfrak{m}}

\def\Lie{\mathcal{L}}

\AtBeginDocument{%
    \newwrite\bibnotes
    \def\bibnotesext{Notes.bib}
    \immediate\openout\bibnotes=\jobname\bibnotesext
    \immediate\write\bibnotes{@CONTROL{REVTEX41Control}}
    \immediate\write\bibnotes{@CONTROL{%
    apsrev41Control,author="08",editor="1",pages="1",title="0",year="1"}}
    \if@filesw
    \immediate\write\@auxout{\string\citation{apsrev41Control}}%
    \fi
}

\begin{document}
\title{Dynamical descalarization in binary black hole mergers}

\begin{abstract}
Scalar fields coupled to the Gauss--Bonnet invariant can undergo a tachyonic
instability, leading to spontaneous scalarization of black holes. Studies of
this effect have so far been restricted to single black hole spacetimes. We
present the first results on dynamical scalarization in head-on collisions and
quasicircular inspirals of black hole binaries with numerical relativity
simulations. We show that black hole binaries can either form a scalarized
remnant or dynamically \emph{descalarize} by shedding off its initial scalar
hair. The observational implications of these findings are discussed.
\end{abstract}

\author{Hector O. Silva}
\email{hector.silva@aei.mpg.de}
\affiliation{Max-Planck-Institut f\"ur Gravitationsphysik (Albert-Einstein-Institut), Am M\"uhlenberg 1, D-14476 Potsdam, Germany}
\affiliation{Illinois  Center  for  Advanced  Studies  of  the  Universe and Department of Physics, University of Illinois at Urbana-Champaign, Urbana, Illinois 61801, USA}

\author{Helvi Witek}
\email{hwitek@illinois.edu}
\affiliation{Illinois  Center  for  Advanced  Studies  of  the  Universe and Department of Physics, University of Illinois at Urbana-Champaign, Urbana, Illinois 61801, USA}

\author{Matthew Elley}
\email{matthew.elley@kcl.ac.uk}
\affiliation{Department of Physics, King's College London, Strand, London, WC2R 2LS, United Kingdom}

\author{Nicol\'as Yunes}
\email{nyunes@illinois.edu}
\affiliation{Illinois  Center  for  Advanced  Studies  of  the  Universe and Department of Physics, University of Illinois at Urbana-Champaign, Urbana, Illinois 61801, USA}

\date{{\today}}

\maketitle

\noindent \bfi{Introduction.}
\vphantom{\bh}
Despite the elegance of Einstein's theory, it presents several shortcomings:
explaining the late-time acceleration of the Universe and providing a consistent
theory of quantum gravity or the presence of spacetime singularities [e.g.,~in
\bhs].
Candidate theories (of quantum gravity) that remedy these shortcomings
typically predict the coupling to additional fields or higher curvature
corrections~\cite{Berti:2015itd}.
Binary \bhs{,} their \gw emission, and the first \gw detections by the LIGO-Virgo Collaboration~\cite{LIGOScientific:2018mvr,LIGOScientific:2020ibl}
offer unique insights into the nonlinear regime of gravity that
unfolds during the \bhs{'} inspiral and merger
and enable new precision tests of gravity~\cite{Yunes:2013dva,Yagi:2016jml}.
So far, these tests have been parametrized null tests against
\gr~\cite{LIGOScientific:2019fpa,LIGOScientific:2020tif} or used a mapping between
these parameters and those of specific theories~\cite{Yunes:2016jcc,Nair:2019iur,Carson:2020rea}.
To do the latter, however, requires \gw predictions in specific theories.

One of the most compelling beyond-\gr theories,
\sGB gravity introduces a dynamical scalar field coupled to the
Gauss--Bonnet invariant.
\sGB gravity emerges in the low-energy limit of quantum gravity
paradigms such as string theory~\cite{Metsaev:1987zx}, through a dimensional
reduction of Lovelock gravity~\cite{Charmousis:2014mia} and is the simplest
model that contains higher curvature operators.
The most studied class of \sGB gravity with a dilatonic or linear coupling to
the scalar field gives rise to hairy
\bhs~\cite{Mignemi:1992nt,Kanti:1995vq,Torii:1996yi,Yunes:2011we,Sotiriou:2013qea,Sotiriou:2014pfa,Prabhu:2018aun}.
This theory, however, has been strongly constrained with \gw
observations from binary \bhs~\cite{Nair:2019iur}.

We turn our attention to another interesting class of \sGB
gravity that is both unconstrained by \gw observations and gives rise to
(spontaneously) scalarized \bhs~\cite{Doneva:2017bvd,Silva:2017uqg}.
Spontaneous scalarization is a familiar concept in beyond-\gr theories;
e.g.,~it is well established for neutron stars in scalar-tensor theories~\cite{Damour:1993hw,Damour:1996ke}.
In such theories, the neutron star matter itself can induce a tachyonic
instability that spontaneously scalarizes the star~\cite{Harada:1997mr}.
When placed in a binary system, initially unscalarized neutron stars can
scalarize dynamically near their merger or a scalarized neutron star can induce
a scalar field in their unscalarized companion~\cite{Barausse:2012da,Palenzuela:2013hsa,Shibata:2013pra,Sennett:2017lcx}.
%
In \sGB gravity, it is the spacetime curvature itself that induces
scalarization of \bhs~\cite{Silva:2017uqg,Doneva:2017bvd},
although this has only been shown for isolated \bhs so far.
%
In this Letter we investigate, for the first time, dynamical scalarization in
{\textit{binary}} \bhs.
We concentrate on head-on collisions of \bhs, but
also present the first binary \bh inspiral study.
Before doing so, it is convenient to first review the basics of \sGB gravity
and spontaneous \bh scalarization.
%

\vspace{0.2cm}
\noindent \bfi{Scalar Gauss--Bonnet gravity and scalarization.}
\sGB gravity is described by the action
\begin{align}
\label{eq:ActionSGB}
S =  \frac{1}{16\pi} \int \dif^{4} x \sqrt{-g} \left[
       \R4D
     - \frac{1}{2} \left( \nabla\Phi \right)^{2}
     + \frac{\aGB}{4} f(\Phi) \, \RGB
     \right]
\,,
\nonumber \\
\end{align}
where a real scalar field $\Phi$ is coupled to the Gauss--Bonnet invariant
$\RGB = \R4D^{2} - 4 \R4D_{\mu\nu} \R4D^{\mu\nu} + \R4D_{\mu\nu\rho\sigma} \R4D^{\mu\nu\rho\sigma}$,
through the function $f(\Phi)$ and a dimensionful coupling constant $\aGB$.
We use geometrical units, $c = 1 = G$, in which $\aGB$ has units of $[{\rm length}]^{2}$.
The action~\eqref{eq:ActionSGB} gives rise to the scalar field equation of motion
\begin{equation}
\label{eq:SGBEoMScalar}
\Box\Phi   = - ({\aGB}/{4}) f'(\Phi) \, \RGB
\,,
\end{equation}
where we defined $(\cdot)' = \dif (\cdot) /\dif\Phi$.
The
function $f(\Phi)$ selects different ``flavors'' of \sGB gravity~\cite{Antoniou:2017acq,Antoniou:2017hxj}.
One subset of these theories has $f'\neq 0$ everywhere.
It includes variants of \sGB gravity with
dilatonic $f(\Phi) \propto \exp(\Phi)$~\cite{Mignemi:1992nt,Kanti:1995vq,Torii:1996yi}
or shift-symmetric $f(\Phi) \propto \Phi$~\cite{Sotiriou:2013qea,Sotiriou:2014pfa,Maselli:2015tta} couplings,
in which \bhs always have scalar hair~\cite{Yagi:2015oca,Prabhu:2018aun}.
Another interesting class of \sGB theories admits an extremum
$f'(\Phi_{0}) = 0$ for a constant $\Phi_{0}$.
They give rise to an effective space-dependent mass term $m^{2}_{\rm eff} = - f''(\Phi_0) \RGB$.
This class includes
quadratic $f(\Phi) \propto \Phi^{2}$~\cite{Silva:2017uqg,Macedo:2019sem}
and Gaussian $f(\Phi) \propto \exp(\Phi^2)$~\cite{Doneva:2017bvd} models.

The latter class still admits all vacuum (\bh) solutions of \gr together with $\Phi=\Phi_{0}=$~const.
In fact, if $f''(\Phi_0) \RGB < 0$ these solutions are unique due to a no-hair theorem~\cite{Silva:2017uqg}.
A linear stability study of these $\Phi_0 = \textrm{const.}$ solutions around a
Schwarzschild \bh reveals that this condition is a requirement for the
absence of a tachyonic instability ($m_{\rm eff}^2 > 0$) for the scalar field
perturbations~\cite{Silva:2017uqg}.
If the effective mass $m^{2}_{\rm eff} < 0$, a tachyonic instability is triggered and the
the \sGB scalar field is excited and {\textit{spontaneously scalarizes}} the \bh{.}
This linear instability~\cite{Macedo:2020tbm} is quenched at the nonlinear level,
resulting in a scalarized \bh as end-state~\cite{Ripley:2020vpk}.
The simplest theory that admits scalarized \bhs is described by the quadratic
coupling
$
f(\Phi) = \bar{\beta}_{2} \, \Phi^{2} \,,
$
where $\bar{\beta}_{2}=$~const.~\cite{note_notation}.
The relevant parameter in this theory is the dimensionless constant
$\beta_{2} = ({\aGB}/{m^2}) \bar{\beta}_{2}$,
where $m$ is the characteristic mass of the system.

The onset of scalarization is fully determined by the scalar's linear dynamics
on a given \gr background.
For a Schwarzschild \bh of mass $m$,
for which $\RGB \geqslant 0$ everywhere,
scalarization first occurs for a spherically symmetric scalar field if
$\beta_{2} = \beta_{\rm c} \sim 1.45123$,
a result in agreement with nonlinear calculations~\cite{Silva:2017uqg,Doneva:2017bvd}.
For values below $\beta_{\rm c}$ the scalar perturbation decays
monotonically at late times (we call them ``subcritical''),
precisely at $\beta_{\rm c}$ the scalar field forms a bound state around the \bh
(``critical''),
and above it the scalar field growths exponentially with time (``supercritical'').
This result was recently generalized to Kerr \bhs, where spin-induced
scalarization can take place for $\beta_2 < 0$, for dimensionless spin
parameters $\chi \geqslant 0.5$~\cite{Dima:2020yac,Hod:2020jjy,Doneva:2020nbb,Doneva:2021dqn}.
Nonlinear rotating scalarized \bh solutions in \sGB gravity were found for both
positive~\cite{Cunha:2019dwb,Collodel:2019kkx} and negative values of
$\beta_2$~\cite{Herdeiro:2020wei,Berti:2020kgk}.
So far studies of scalarization in \sGB gravity focused on single \bhs.
We advance these studies to \bh binaries, and expand upon~\cite{Witek:2018dmd},
focusing on the quadratic theory $f(\Phi) = \bar{\beta}_2 \Phi^2$, as discussed next.

\vspace{0.2cm}
\noindent \bfi{Numerical methods and simulations.}
%
We investigate \bh scalarization in the decoupling limit, i.e., we numerically
evolve the scalar field on a time-dependent background in vacuum \gr that
represents binary \bh spacetimes.
Unless stated otherwise, we follow the approach of~\cite{Witek:2018dmd}
and refer to it for details.
We foliate the spacetime into spatial hypersurfaces with $3$-metric $\gamma_{ij}$ and
extrinsic curvature
$K_{ij} = - (2 \alpha)^{-1} \, \dif_{t} \gamma_{ij}$,
where $\dif_{t} = \partial_{t} - \Lie_{\beta}$, $\Lie_{\beta}$ being
the Lie derivative along the shift vector $\beta^{i}$, and $\alpha$ is the lapse function.
We write Einstein's equations as a Cauchy problem
and adopt the \bssn formulation~\cite{Shibata:1995we,Baumgarte:1998te} of the time evolution equations
complemented with the moving-puncture gauge conditions~\cite{Campanelli:2005dd,Baker:2005vv}.
%
We prepare
Brill-Lindquist initial data~\cite{Brill:1963yv,Lindquist:1963jmp}
for head-on collisions or Bowen-York initial data~\cite{Bowen:1980yu,Brandt:1997tf} for
a quasicircular \bh binary.

To evolve the scalar field, we introduce its momentum $\Kphi = - \alpha^{-1} \, \dif_{t} \Phi$,
and write its field equation~\eqref{eq:SGBEoMScalar} as
\begin{align}
\label{eq:EvolEquationsPhiKphi}
\dif_{t} \Phi  & = - \alpha\Kphi
\,,\\
\dif_{t} \Kphi & = - D^{i}\alpha D_{i}\Phi
               - \alpha\left( D^{i}D_{i}\Phi - K\Kphi + \frac{\aGB}{4} f' \RGB \right)
\,,\nonumber
\end{align}
where $D_i$ is the covariant derivative associated with $\gamma_{ij}$,
$K = \gamma^{ij}K_{ij}$,
$f'= 2 \bar{\beta}_2 \,\Phi$,
and $\RGB$ is the Gauss--Bonnet invariant of the background spacetime.
We set the system's total mass to unit, i.e., $M = m_1 + m_2 = 1$, where
$m = m_{1,2}$ is the component's mass.
%
The scalar field is initialized either as a spherically symmetric Gaussian
shell (G) located at $r_{0} = 12M$ and with width $\sigma = 1 M$ as in~\cite{Witek:2018dmd} or as a bound
state (B) around each binary component,
\begin{align}
\label{eq:SFInitData}
    \Phi|_{t=0} &= 0\,,
    \,\,\, K_{\Phi}|_{t=0} = \frac{1}{\sqrt{4\pi}} \exp\left[\frac{(r - r_0)^2}{\sigma^2}\right]\,,
    \\
    \Phi|_{t=0} &=
    \frac{m r}{\varrho^2} \left[
    c_1
    + \frac{c_2 m r}{\varrho^2}
    + \frac{c_3 (m r)^2}{\varrho^4}
    \right]\,, \,\,\, \Kphi|_{t=0} = 0\,.\nonumber
\end{align}
Here, $\varrho = m+2r$, and $c_1 = 3.68375$, $c_2 = 4.972416$, $c_3 = 4.972416 \cdot 10^{2}$ are
fitting constants to reproduce the numerical results in~\cite{Silva:2017uqg}.

\begin{figure}[t]
\begin{tabular}{cc}
    \includegraphics[width=.325\columnwidth]{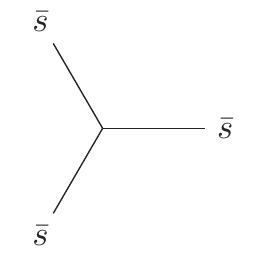} &
    \includegraphics[width=.325\columnwidth]{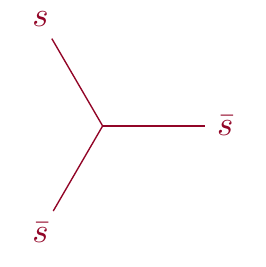}
    \\
    {(a) -- $\{ {\rm G}, \, 1,   \, 0 \}$} &
    {(b) -- $\{ {\rm B}, \, 1/2, \, 0.16125 \}$}
    \\ \\
    \includegraphics[width=.325\columnwidth]{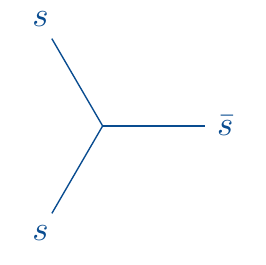} &
    \includegraphics[width=.325\columnwidth]{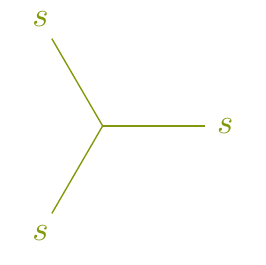}
    \\
    {(c) -- $\{ {\rm B}, \, 1, \, 0.36281 \}$} &
    {(d) -- $\{ {\rm B}, \, 1, \, 1.45123 \}$}
\end{tabular}
\caption{\label{fig:diagram}
Summary of simulations of \bh head-on collisions,
where $\bar{s}$ and $s$ stand for initial or final states
that are either nonscalarized or scalarized respectively.
Each diagram is labeled by the initial data
(Gaussian shell ``G'' or bound state ``B''),
the mass ratio $q = m_1/m_2$ ($1$ or $1/2$)
and the coupling parameter $\beta_2$.
In case (a) (top left panel) two nonscalarized BHs produce a nonscalarized
remnant. In case (b) (top right panel) a scalarized and a nonscalarized BH
produce a nonscalarized remnant.  This initial configuration is possible when $q$
is different from one. In case (c) (bottom left panel) two scalarized BHs
produce a nonscalarized remnant. Finally, in case (d) (bottom right panel) two
scalarized BHs produce a scalarized remnant.
}
\end{figure}

We perform our numerical simulations with \canuda~\cite{witek_helvi_2020_3565475,
Benkel:2016kcq,Benkel:2016rlz,Witek:2018dmd},
coupled to the open-source \ETK~\cite{steven_r_brandt_2020_3866075,Loffler:2011ay}.
We extended the implementation of~\cite{Benkel:2016kcq,Benkel:2016rlz,Witek:2018dmd}
to general coupling functions $f$, including the quadratic coupling.
We employ the method of lines with fourth-order finite difference stencils to realize spatial derivatives and a fourth-order Runge-Kutta time integrator.
We use box-in-box mesh refinement provided by {\textsc{Carpet}}~\cite{Schnetter:2003rb}.
The numerical grid contains seven refinement levels, with the outer boundary located at $256M$
and a grid spacing of $\dif x = 1.0M$ on the outer mesh.
To assess the numerical accuracy of our simulations we evolved case (b) in Fig.~\ref{fig:diagram}
with additional resolutions $\dif x = 0.9M$ and $\dif x = 0.8M$.
We find second-order convergence and a relative discretization error of $\Delta\Phi_{00}/\Phi_{00} \lesssim0.5\%$,
where $\Phi_{00}$ is the $\ell = \mathfrak{m} = 0$ multipole of the scalar field.
We present the corresponding convergence plot for the scalar monopole and for
the gravitational wave $\ell = 2$, $\mathfrak{m} = 0$ mode in Fig.~\ref{fig:Phi00ConvergencePlot} of the
Supplemental Material.

\vspace{0.2cm}
\noindent \bfi{Results.}
%
We performed a large set of \bh head-on collisions with varying mass ratio
$q=m_{1}/m_{2} \leqslant 1$, total mass $M = 1$
and initial separation $d = 25M$, considering both initial data in Eq.~\eqref{eq:SFInitData}.
The \bhs merge at $t_{\rm M} \sim 179.5 M$, as estimated from the peak of the $\ell=2,\mmode=0$ multipole
of the gravitational waveform.
To guide our choices of $\beta_2$,
we recall that the critical coupling for the fundamental mode is
$\beta_{2,\rm c} = \beta_{\rm c} \, (m / M)^2$ with
$\beta_{\rm c} \sim 1.45123$,
and $m$ denotes either the individual \bhs' mass $m_{1,2}$
or the total mass $M$.
For example, for an equal-mass binary with $m_{1}=m_{2}=M/2$, the critical coupling
for the individual holes is
$\beta^{(1)}_{2,\rm c} = \beta^{(2)}_{2,\rm c}  = \beta_{\rm c} / 4 = 0.36275$
and that of the final hole is approximately
$\beta^{\rm f}_{2,\rm c} = \beta_{\rm c}$ where
we neglected the small mass loss in the form of \gw{s} during the
collision~\cite{Lousto:2004pr,Sperhake:2011ik}.

Here we present a selection of our results, illustrated in
Fig.~\ref{fig:diagram}, to highlight our most important findings.
An expanded discussion will be presented in a companion
paper~\cite{Elley:2020inspiralprd}.
We vary the initial state by setting the coupling parameter $\beta_{2}$ such that
\begin{enumerate*}[label={(\alph*)}]
\item none of the \bhs are initially scalarized,
\item the smaller-mass \bh initially carries a bound-state scalar field,
both \bhs carry initially a bound-state scalar that leads either to a nonscalarized
final \bh [case \item{}\!\!] or a scalarized final \bh [case \item{}\!\!].
\end{enumerate*}

\begin{figure}[t]
\includegraphics[width=\columnwidth]{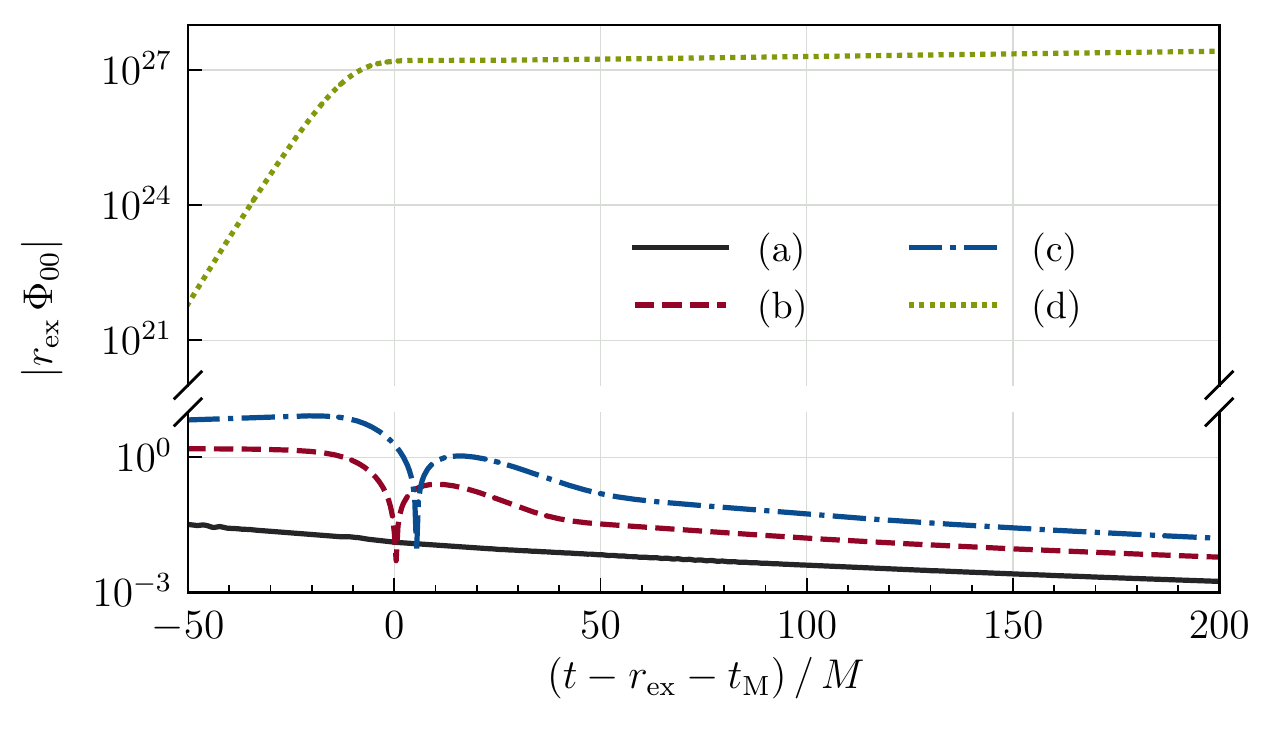}
\caption{\label{fig:Phi00Evolution}
Time evolution of the scalar field $\ell = \mmode = 0$ multipole in the
background of a \bh head-on collision with initial separation $d = 25M$.
It is rescaled by the extraction radius $\rex=50M$ and shifted in time
such that $(t-\rex-t_{\rm M})/M=0$ corresponds to the \bhs{'} merger.
The labels refer to the four cases summarized in Fig.~\ref{fig:diagram}.
}
\end{figure}

In Fig.~\ref{fig:Phi00Evolution} we show the $\ell=\mmode=0$ scalar field
multipole extracted on a sphere of fixed radius $\rex=50M$, as a function of
time, and we present snapshots of the scalar's profile in
the Supplemental Material.
%
In case (a), the scalar perturbation is not supported at all (since $m_{\rm eff} = 0$)
and, indeed, after a brief interaction at early times
it decays already before the \bhs collide.
%
In cases (b) and (c) we find a constant scalar field before the \bhs collide,
that is consistent with a bound state around the individual ($q=1$) or smaller-mass \bh ($q=1/2$).
After the merger the scalar field decays since the curvature (and thus $m_{\rm eff}$) decreases
and the system no longer supports a bound state -- the final \bh dynamically
{\emph{descalarizes}}.
%
In case (d), the scalar field grows exponentially before the merger because it
is supercritical for the individual \bhs and settles to a constant in time
that is consistent with a bound state around the final \bh.

\begin{figure*}[ht]
\includegraphics[width=\textwidth]{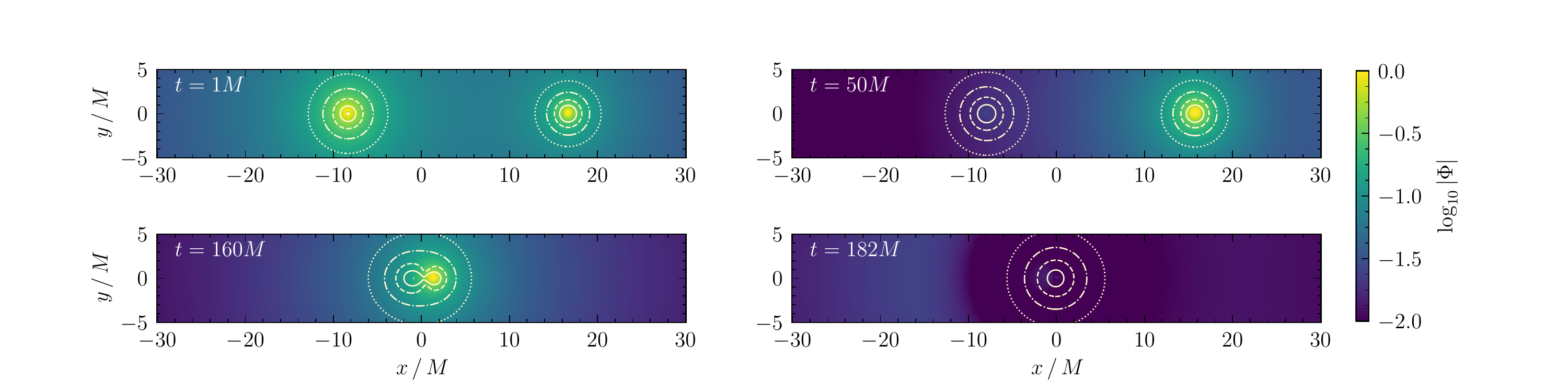}
\caption{
Scalar field and Gauss--Bonnet dynamics on the $xy$--plane for case (b).
We show the amplitude of $\log_{10}|\Phi|$ (color map) together with the
Gauss--Bonnet invariant (isocurvature levels) at the beginning of the evolution
(top left), during the \bhs{'} approach (top right), shortly before the
collision (bottom left) and shortly after the merger (bottom right).
The isocurvature levels correspond to $1 M^{-4}$ (solid
line), $10^{-1} M^{-4}$ (dashed line), $10^{-2} M^{-4}$ (dot-dashed line) and
$10^{-3} M^{-4}$ (dotted line).
}
\label{fig:xy_snapshots}
\end{figure*}

In Fig.~\ref{fig:xy_snapshots} we show two-dimensional snapshots of the scalar
field and spacetime curvature for case (b) which
illustrates the dynamical descalarization phenomenon~\cite{note_web}.
The color map is shared among all panels and shows the amplitude of
$\log_{10}|\Phi|$, while the curves
are isocurvature levels
of $\mathscr{G} M^{4} = \{ 1, 10^{-1}, 10^{-2}, 10^{-3}\}$.
Initially, at $t=1M$, both \bhs
(whose locations are revealed by the isocurvature levels) are surrounded by nontrivial scalar field
profiles given by Eq.~\eqref{eq:SFInitData}.
At $t=50M$, the smaller \bh hosts a bound state scalar that is dragged along the hole's motion,
inducing scalar dipole radiation that would impact the GWs emitted.
In contrast, the scalar field around the larger \bh
disperses because its curvature is too small to sustain a bound state for
a coupling of $\beta_{2}=0.36281$.
The system thus evolves as a $s + \bar{s}$ process in the notation of Fig.~\ref{fig:diagram}.
At $t=160M$, the \bhs are about to merge, as indicated by the two lobes in the isocurvature contours,
the curvature of the combined system decreases and the scalar field starts dissipating.
At $t=182M$, which is shortly after the collision, the system has descalarized
since for the final \bh $\beta^{\rm f}_{2, {\rm c}} > \beta_2$.

\begin{figure}
\includegraphics[width=\columnwidth]{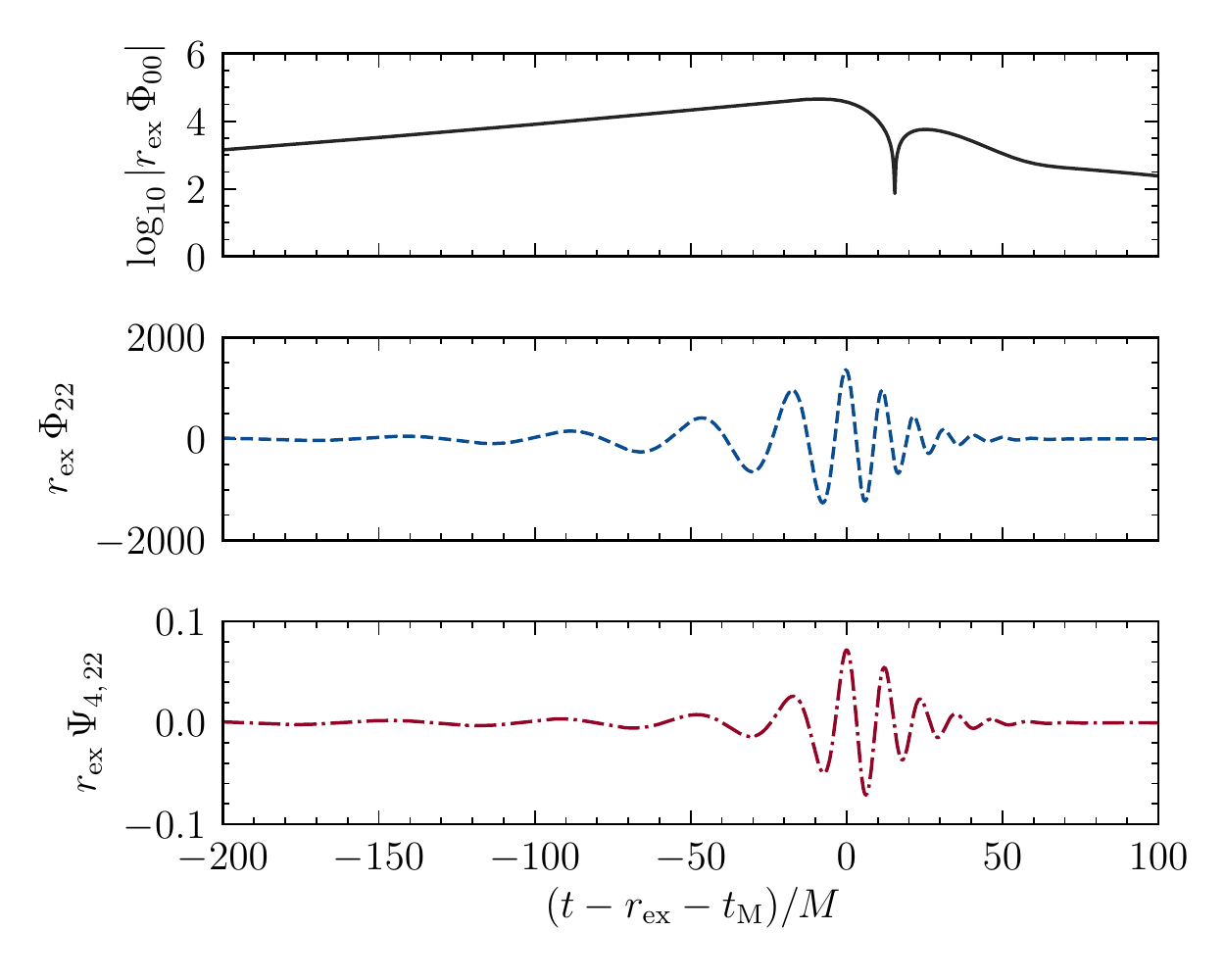}
\caption{Scalar and gravitational waveforms, rescaled by the extraction radius $r_{\rm ex}=50M$,
sourced by an equal-mass \bh binary with bound state initial data on each \bh. This system is
the inspiral counterpart of case (c) and shows dynamical descalarization in action.}
\label{fig:inspiral_waveforms}
\end{figure}

We also simulated the inspiral of an equal-mass, nonspinning \bh binary
with initial separation of $d=10M$, $\beta_2 = 0.36281$, and bound state scalar field
initial data.
%
This corresponds to an initial configuration in which both \bhs are scalarized,
and then, after merger, the remnant is not scalarized, which is
analogous to case (c) of Fig.~\ref{fig:diagram} in the head-on case.
In Fig.~\ref{fig:inspiral_waveforms}, we show the gravitational quadrupole waveform (bottom panel),
as characterized by the  $\ell=\mathfrak{m}=2$ mode of the Newman-Penrose scalar $\Psi_{4}$,
together with the scalar field's monopole (top) and quadrupole (middle).
The scalar's monopole $\Phi_{00}$ exhibits the distinctive signature of
descalarization:
the increase in the field's amplitude during the inspiral of scalarized \bhs
is followed by a complete dissipation of the scalar field after the merger ($t_{\rm M} \sim 917 M$) as the curvature of the remnant \bh
no longer supports a bound state.
In addition, the dynamics of the \bh binary sources scalar quadrupole radiation
(of the initially spherically symmetric scalar).
The field's amplitude grows exponentially during the inspiral
and decays after the \bhs have merged.
The origin of this excitation is not direct scalarization of the $\ell=2$ scalar bound state, but due to the inspiral of
two scalarized (or ``hairy'') \bhs{.}
This interpretation is further supported by the observation that the phase
of the $\ell=\mathfrak{m}=2$ scalar mode
is driven by the binary's orbital frequency.
We also observed this for the $\ell = \mathfrak{m} = 4$ mode and expect it to
happen for all even $\ell = \mathfrak{m}$ modes.
For $q=1$, the odd $\ell = \mathfrak{m}$ modes are suppressed due to symmetry,
whereas they would be excited in the general case $q \neq 1$.
The descalarization during the merger is reminiscent of the decrease in scalar
charge observed in the shift-symmetric theory~\cite{Witek:2018dmd}, however,
with the striking difference that here the remnant \bh is a rotating \gr
solution.

\vspace{0.2cm}
\noindent \bfi{Discussions.}
We presented the first numerical relativity simulations of the scalar field
dynamics in binary \bh spacetimes in quadratic \sGB gravity~\cite{Silva:2017uqg}.
We found that the interplay between mass ratio $q$ and $\beta_{2}$ can result
in different scenarios for the scalar field dynamics. Most notably, it can
lead to a {\textit{dynamical descalarization}} of the binary, which we observed in both
head-on and quasicircular inspiral simulations.
%
Here we focused on $\beta_2 \geqslant 0$, but the case $\beta_2 < 0$ would be
particularly interesting to study in inspiral simulations. More specifically,
the spinning remnant of a binary \bh merger
typically has a dimensionless spin
$\chi \sim 0.7$~\cite{Buonanno:2006ui}, sufficient to trigger a spin-induced
tachyonic instability of the scalar field~\cite{Dima:2020yac}.
This is currently under study~\cite{Elley:2020inspiralprd}. It would be
interesting to frame this effect within the \eft of~\cite{Khalil:2019wyy} or
in a post-Newtonian framework~\cite{Yagi:2011xp,Julie:2019sab,Shiralilou:2020gah},
although the latter may not be suitable for the modeling of a nonlinear
dynamical scalarization process.

The scalar excitations we have discovered during binary \bh coalescence in this class of sGB theories
have important implications to GW observations and tests of GR. In particular, the scalar excitations
will drain the binary of energy as they propagate away from the system, the monopole scalar piece inducing dipole losses, and the
quadrupole piece correcting the quadrupole \gw losses of \gr, which, based on~\cite{Wagle:2019mdq}, are
expected to only have the same ``plus'' and ``cross'' polarizations.
This enhanced dissipation of energy and angular momentum, in turn, will force
the binary to inspiral faster than in \gr, and therefore, leave an imprint in
the \gw{s} emitted through corrections to the rate at which the GW frequency
increases during the inspiral.
This GW phase shift will enable us to project bounds on sGB gravity that are
similar in spirit but complementary to the analysis of~\cite{Witek:2018dmd}.
%
In fact, because the merger leaves behind a ``bald'' Kerr black hole due to
dynamical descalarization, the (scalar) energy flux is, in general, larger as
compared to shift-symmetric sGB, where the remnant black hole always retains
some of its hair.
This suggests that strong observational bounds might be placed on this theory.

Having worked in the decoupling limit, a question naturally arises: what would we
expect in the fully nonlinear regime of \sGB gravity?
It is known that nonlinear effects set an upper bound on the scalar field magnitude
at the \bh horizon~\cite{Antoniou:2017acq}, so that the domain of existence of
scalarized \bhs exhibits a very narrow bandlike structure in the phase space spanned by
\bh mass and coupling $\beta_{2}$; see Fig.~2 of~\cite{Silva:2017uqg}.
This means that case (d) would only occur for sufficiently small mass ratios such that both
the initial binary and its final state remain in band.
In general, however, comparable mass \bh binaries could undergo an
$\bar{s} + \bar{s} \to s$ process, in which two
unscalarized \bhs would merge, forming \bh within the scalarization band, i.e.,~a dynamical \bh scalarization.
The descalarization of the \bh remnant would also impact the \gw
emission during the ringdown.
Specifically, the waveforms in Fig.~\ref{fig:inspiral_waveforms} show that the
ringdown time scales of scalar and tensorial modes are comparable. This
suggests that
one should expect to see the imprint of the descalarization onto the
quasinormal mode spectra of the Kerr black hole in the nonlinear case.
Performing these studies in practice would require a general, well-posed
formulation of the time evolution equations outside the \eft
approach~\cite{Witek:2018dmd,Okounkova:2020rqw}, small values of the coupling
parameter~\cite{Kovacs:2020ywu,Kovacs:2020pns}, or
spherical symmetry~\cite{Ripley:2019hxt,Ripley:2019irj,Ripley:2019aqj,Ripley:2020vpk}.
Finding such a formulation has proven
challenging~\cite{Papallo:2017qvl,Papallo:2017ddx,Julie:2020vov,Witek:2020uzz},
although first results in this direction were presented in~\cite{East:2020hgw}.
Our work motivates and paves the way for future studies of nonperturbative,
beyond-\gr effects in \bh binaries, with potential implications to tests of \gr
with \gw astronomy.

\vspace{0.2cm}
\noindent \bfi{Acknowledgments.}
%
We thank Katy Clough, Mohammed Khalil, and Jan Steinhoff for useful discussions.
%
H.W.~acknowledges financial support provided
by
the NSF Grant No.~OAC-2004879,
the Royal Society University Research Fellowship Grant No.~UF160547,
and
the Royal Society Research Grant No.~RGF\textbackslash R1\textbackslash 180073.
H.O.S and N.Y.~acknowledge financial support through NSF Grants
No. PHY-1759615 and PHY-1949838, and NASA ATP Grants No.~17-ATP17-0225, No.~NNX16AB98G and
No.~80NSSC17M0041.
%
We thankfully acknowledge the computer resources and the technical support
provided by the Leibniz Supercomputing Center via PRACE Grant No.~2018194669
``FunPhysGW: Fundamental physics in the era of gravitational waves'' and by the
DiRAC Consortium via STFC DiRAC Grants No.~ACTP186 and No.~ACSP218.
This work made use of the Illinois Campus Cluster, a computing resource that is
operated by the Illinois Campus Cluster Program (ICCP) in conjunction with the
National Center for Supercomputing Applications (NCSA) and which is supported
by funds from the University of Illinois at Urbana-Champaign.

\bibliography{../master_biblio}

\clearpage
\newpage
\onecolumngrid
\FloatBarrier

\section*{-- Supplemental Material --}

\section{Convergence plots}

\begin{figure}[ht]
\includegraphics[width=0.45\textwidth]{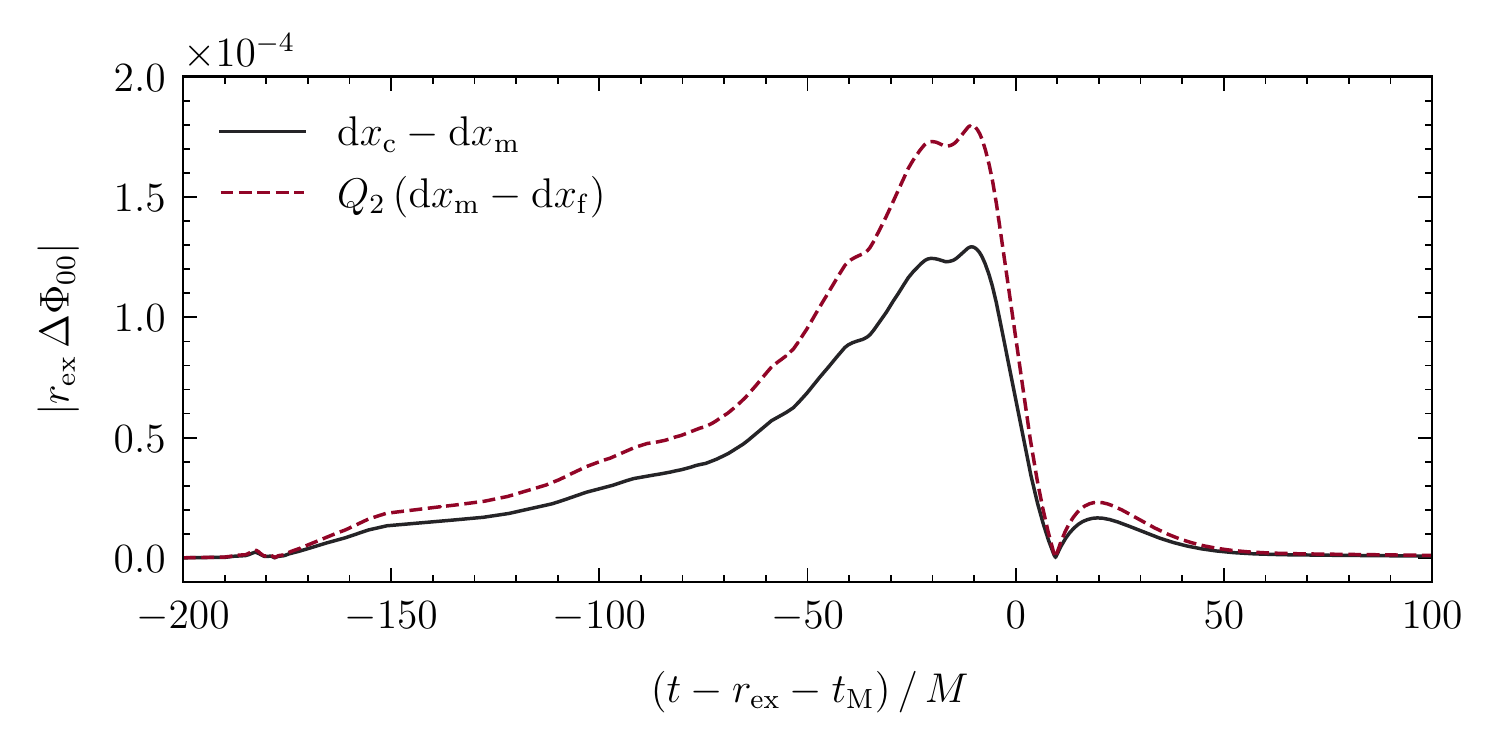}
\includegraphics[width=0.45\textwidth]{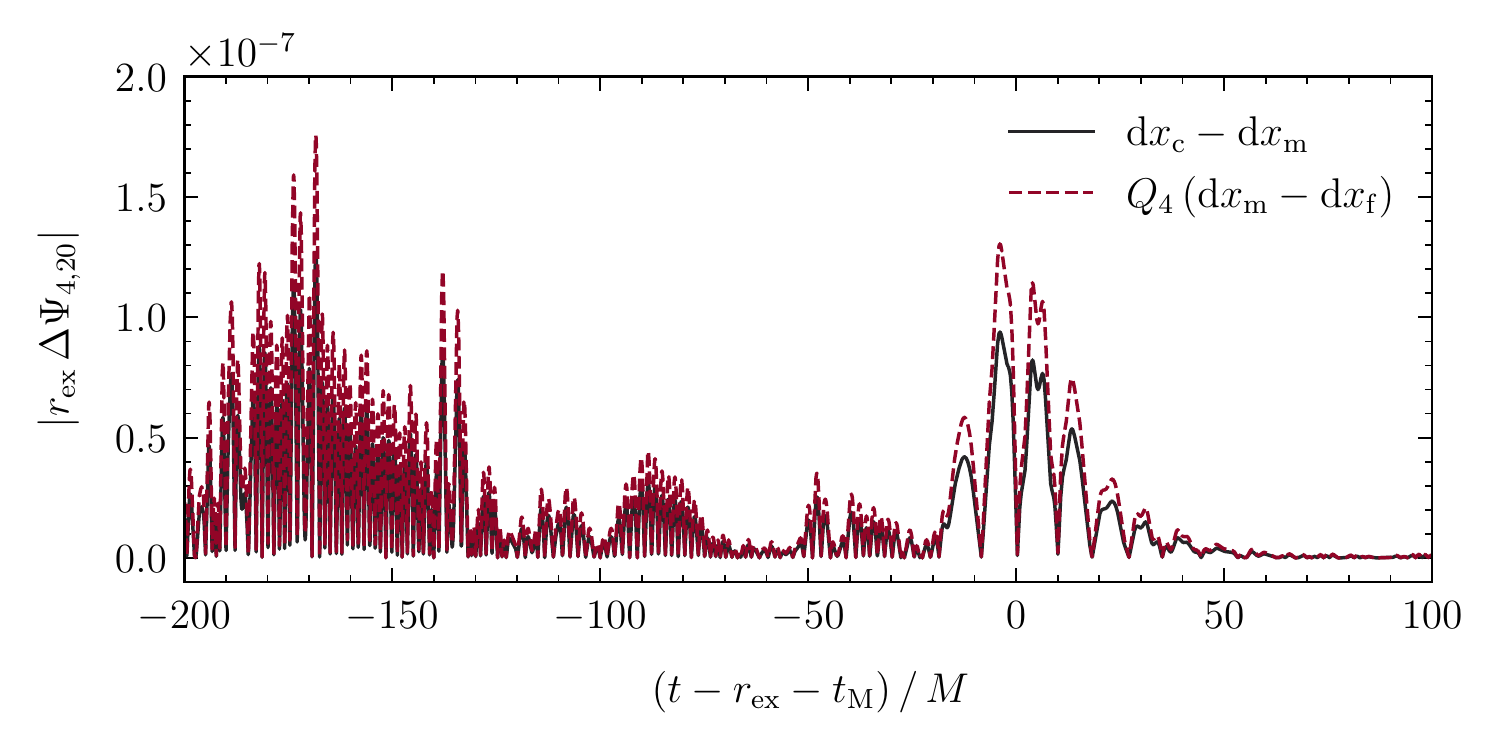}
\caption{\label{fig:Phi00ConvergencePlot}
{
{Convergence plot for case (b) in Fig.~\ref{fig:diagram} of the main text.
We show the scalar field monopole (left panel) and the gravitational quadrupole
(right panel) extracted at $\rex=100M$ and shifted in time such that $(t - r_{\rm ex} - t_{\rm M}) / M = 0$
corresponds to the BHs' merger.
We calculate the differences between the coarse and medium resolutions, $\dif x_{\rm c}=1.0M$ and $\dif x_{\rm m}=0.9M$ (solid line),
and medium and high resolutions, $\dif x_{\rm m}=0.9M$ and $\dif x_{\rm f}=0.8M$.
For the scalar field monopole we rescale the latter by $Q_{2}=1.12$ (dashed line), indicating second-order convergence.
For the gravitational quadrupole we rescale by $Q_4 = 1.39$ (dashed line), indicating fourth-order convergence, as stated
in the main text.
}
}
}
\end{figure}

{We assess the discretization error of our simulations by exemplarily running the head-on collision of
equal-mass black holes that initially carry a bound-state scalar field with coupling parameter $\beta_{2}=0.36281$
at three different resolutions $\dif x_{\rm c}=1.0M$, $\dif x_{\rm m}=0.9M$ and $\dif x_{\rm f}=0.8M$.
Here $M$ is the system's total mass, which we set to unit.
This setup corresponds to case (c) in Fig.~\ref{fig:diagram} of the main text.
Focusing on the scalar field monopole ($\Phi_{00}$) and the gravitational quadrupole ($\Psi_{4,20}$) we
compute the differences between the course and medium, and medium and high resolution runs.
For $\Phi_{00}$, we rescaled the latter difference by the convergence factor $Q_{2}=1.12$, as shown in
the left panel in Fig.~\ref{fig:Phi00ConvergencePlot}, indicating second-order convergence.
For $\Psi_{4,20}$, we rescaled the latter difference by $Q_{4}=1.39$, as shown in
the right panel, indicating fourth-order convergence.
Computing the relative difference $\Delta \Phi_{00} / \Phi_{00}$ between the coarsest resolution simulation with $\dif x_{\rm c}=1.0M$ and the second-order Richardson extrapolation, we find a numerical error of $\Delta\Phi_{00}/\Phi_{00}\lesssim0.5\%$ as stated in the main text.}


\section{Snapshots of scalar field profile}

\begin{figure*}[htb]
\includegraphics[width=\textwidth]{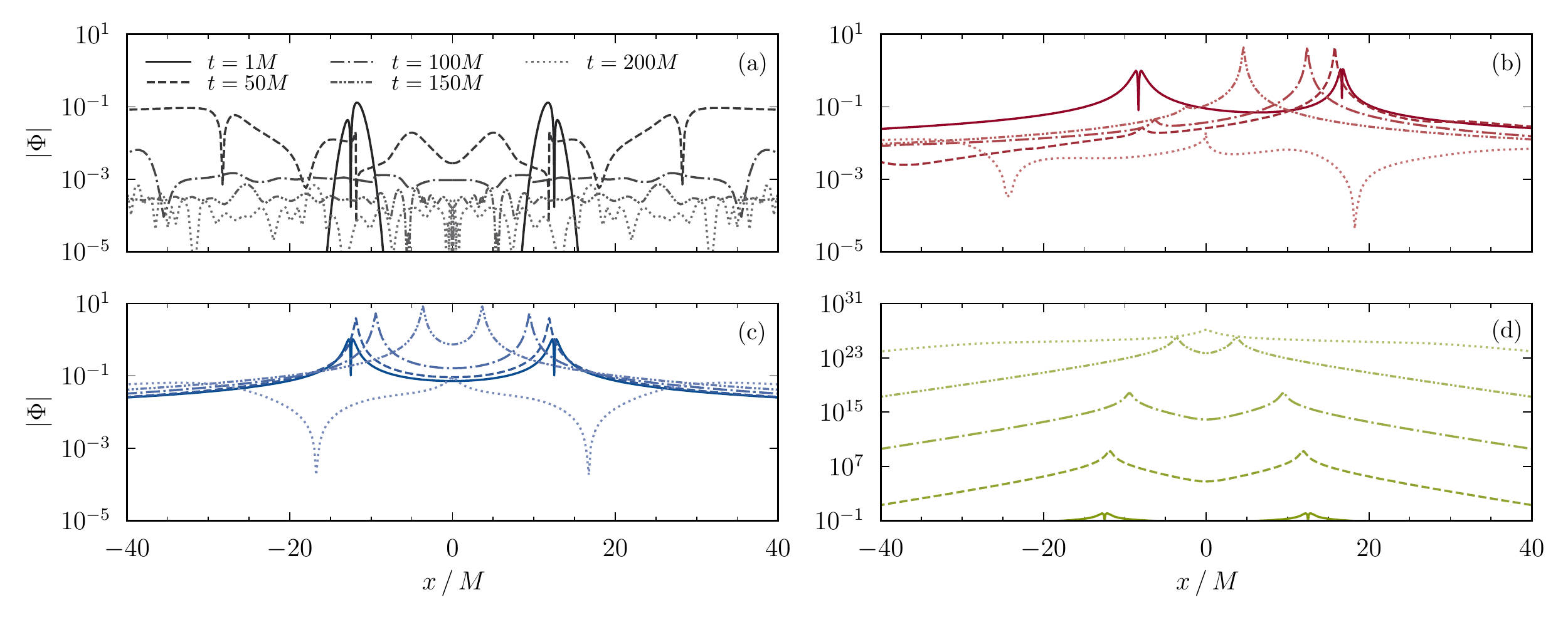}
\caption{\label{fig:PhiSnapshotsAxis}
Scalar field's profile along the collision axis $x/M$ at different instances in time
before, during and after the \bh head-on collision for cases
(a)--(d) defined in Fig.~\ref{fig:diagram}. The merger happens at $t_{\rm M} \sim 179.5M$.
}
\end{figure*}

Figure~\ref{fig:PhiSnapshotsAxis} presents the scalar field profile along the
collision axis $x/M$ at different instances throughout the evolution before,
near and after the merger of the \bhs{.}
In case (a), the scalar field is below the critical value to form any bound
state configurations and, indeed, after a brief interaction at early times
it decays already before the \bhs collide.
In cases (b) and (c), the scalar field forms a bound state that is anchored
around the individual ($q=1$) or smaller-mass \bh ($q=1/2$). As the \bhs
approach each other, the scalar field follows their dynamics and moves along
the collision course with only small adjustments to its spatial configuration.
After the \bhs merge, the critical value $\beta_{2,{\rm c}}$ to form a bound state
increases, i.e., the \bh can no longer support a scalar bound state.
Consequently, the configuration becomes subcritical and the scalar field is
depleted, indicating dynamical descalarization of the \bh binary.
Finally, case (d) is set up such that the final configuration is near critical
to form a bound state, always leading to a supercritical setup before merger.
Indeed, we observe that the scalar field grows (exponentially), before settling
to a constant-in-time radial profile after the merger.  This rapid growth is
due to the fact that $\beta_2 \sim 1.45123$ is four times larger than the critical
scalarization value for the initial \bhs{.}

\FloatBarrier

\end{document}